\documentclass[journal,twoside,web]{ieeecolor}

\usepackage{tmi}
\usepackage{cite}
\usepackage{amsmath,amssymb,amsfonts}
\usepackage{algorithmic}
\usepackage{graphicx}
\usepackage{textcomp}
\usepackage{booktabs}
\usepackage[colorlinks,linkcolor=blue]{hyperref}
\usepackage{multirow}
\usepackage{bm}
\usepackage{array}
\usepackage{epstopdf} 

\def\ie{\emph{i.e.}}
\def\eg{\emph{e.g.}}
\def\etal{{\em et al.}}
\def\etc{{\em etc.}}

\newcommand{\PreserveBackslash}[1]{\let\temp=\\#1\let\\=\temp}
\newcolumntype{C}[1]{>{\PreserveBackslash\centering}m{#1}}
\newcolumntype{?}[1]{!{\vrule width #1}}

\def\BibTeX{{\rm B\kern-.05em{\sc i\kern-.025em b}\kern-.08emT\kern-.1667em\lower.7ex\hbox{E}\kern-.125emX}}
\begin{document}
\title{Dual-Teacher++: Exploiting Intra-domain and Inter-domain Knowledge with Reliable Transfer for Cardiac Segmentation}
\author{Kang Li, Shujun Wang,~\IEEEmembership{Student Member, IEEE}, Lequan Yu,~\IEEEmembership{Member, IEEE},\\ Pheng-Ann Heng,~\IEEEmembership{Senior Member, IEEE}
\thanks{The work described in this paper was supported by Key-Area
Research and Development Program of Guangdong Province, China under Project
No. 2020B010165004, Hong Kong Innovation and Technology Fund under Project No.
ITS/311/18FP and National Natural Science Foundation of China under Project No. U1813204.}
\thanks{K. Li, S. Wang and P.-A. Heng are with the Department of Computer
Science and Engineering, The Chinese University of Hong Kong,
Hong Kong, China (e-mail: kli@cse.cuhk.edu.hk; sjwang@cse.cuhk.edu.hk;
pheng@cse.cuhk.edu.hk). Pheng Ann Heng is also with the Guangdong Provincial Key Laboratory of Computer Vision and Virtual Reality Technology, Shenzhen Institutes of Advanced Technology, Chinese Academy of Sciences, Shenzhen, China. L. Yu is with the Department of Radiation Oncology, Stanford University, Palo Alto, CA 94306, USA (e-mail: lequany@stanford.edu). L. Yu is the corresponding author of this work.}
}
\maketitle
\IEEEpeerreviewmaketitle

\begin{abstract}
    Annotation scarcity is a long-standing problem in medical image analysis area.
    To efficiently leverage limited annotations,
    abundant unlabeled data are additionally exploited in semi-supervised learning, while well-established cross-modality data are investigated in domain adaptation.
    In this paper, we aim to explore the feasibility of concurrently leveraging both unlabeled data and cross-modality data for annotation-efficient cardiac segmentation.
    To this end, we propose a cutting-edge semi-supervised domain adaptation framework, namely Dual-Teacher++.
    Besides directly learning from limited labeled target domain data (\eg, CT) via a student model adopted by previous literature, we design novel dual teacher models, including an inter-domain teacher model to explore cross-modality priors from source domain (\eg, MR) and an intra-domain teacher model to investigate the knowledge beneath unlabeled target domain.
    In this way, the dual teacher models would transfer acquired inter- and intra-domain knowledge to the student model for further integration and exploitation.
    Moreover, to encourage reliable dual-domain knowledge transfer, we enhance the inter-domain knowledge transfer on the samples with higher similarity to target domain after appearance alignment, and also strengthen intra-domain knowledge transfer of unlabeled target data with higher prediction confidence.
    In this way, the student model can obtain reliable dual-domain knowledge and yield improved performance on target domain data.
    We extensively evaluated the feasibility of our method on the MM-WHS 2017 challenge dataset. The experiments have demonstrated the superiority of our framework over other semi-supervised learning and domain adaptation methods.
    Moreover, our performance gains could be yielded in bidirections,~\ie, adapting from MR to CT, and from CT to MR. Our code will be available at  \href{https://github.com/kli-lalala/Dual-Teacher-}{https://github.com/kli-lalala/Dual-Teacher-.}

\end{abstract}

\begin{IEEEkeywords}
Semi-supervised domain adaptation, cross-modality, cardiac segmentation
\end{IEEEkeywords}

\section{Introduction}
In the past several years, cardiovascular diseases have become the number one cause of death globally and the mortality rate keeps increasing annually\footnote{\url{https://www.who.int/cardiovascular\_diseases/about\_cvd/en/}}.
During the cardiovascular disease diagnosis, image segmentation is capable to partition clinically significant cardiac substructures, which is a commonly used prerequisite task to quantify morphological and pathological changes in human heart~\cite{zhuang2019evaluation}.
The segmented heart substructure regions would provide assistance to doctors in many aspects like planning and monitoring the treatment~\cite{valindria2018multi} and robot-assisted minimally invasive surgery~\cite{jin2019incorporating}.
Contemporary clinical practice always requires multiple imaging modalities for accurate diagnosis, since each modality exhibits distinct properties and image contrast.
In this circumstance, recent image segmentation approaches were proposed to combine the merits of multiple modality data for comprehensive segmentation~\cite{valindria2018multi,dou2020unpaired,chen2019synergistic,li2020towards,zhuang2019evaluation}.

Despite recent advances of deep convolutional neural networks in heart structure segmentation~\cite{vigneault2018omega}, disease classification~\cite{zheng2019explainable,wolterink2017automatic,liu2018vessel}, coronary artery plaque detection~\cite{zreik2018recurrent}, \etc, these successes rely heavily on massive annotated datasets.
However, collecting and labeling such large-scaled dataset is prohibitively time-consuming and expensive, especially in medical image analysis area, where medical annotations require demanding diagnostic expertise~\cite{litjens2017survey}.
To alleviate annotation scarcity, considerable efforts have been devoted to exploit extra information from related data resources.
Among them, semi-supervised learning and domain adaptation are two widely studied learning approaches.

Semi-supervised learning (SSL) aims to leverage unlabeled data to reduce the usage of manual annotations~\cite{lee2013pseudo,laine2016temporal,tarvainen2017mean}.
By utilizing abundant unlabeled data, the model generalization ability could be greatly enhanced.
Meanwhile, cross-modality medical data is widely available~\cite{li2020towards}, as modern clinical practices often utilize multiple imaging modalities of the same anatomy for a comprehensive view in diagnosis.
Plentiful efforts have devoted on domain adaptation (DA) to leverage the prior knowledge of other modalities for enhanced segmentation performance~\cite{ghafoorian2017transfer,perone2019unsupervised,jiang2018tumor,huo2018synseg}.
Among them, multi-modality learning (MML) exploits the labeled data from a related modality (\ie, source domain) to facilitate the analysis on the modality of interest (\ie, target domain)~\cite{van2018learning, valindria2018multi}.
Since multi-modality learning still requires annotations of two modality data, unsupervised domain adaptation (UDA) extends it with a broader application potential~\cite{orbes2019knowledge, dou2018unsupervised,chen2019synergistic}.
In UDA setting, the source domain annotations are still required while none target domain annotation is needed.

The approaches mentioned above have exhibit promising performance in various cardiac applications.
However, semi-supervised learning simply concentrates on leveraging the unlabeled data affiliated to the same domain as labeled ones, ignoring rich prior knowledge (\eg, shape priors) cross modalities.
While domain adaptation can utilize cross-modality priors, it still has considerable space for improvements.
These observations motivate us to explore the feasibility of integrating the merits of both semi-supervised learning and domain adaptation to mitigate annotation scarcity.
Specifically, we aim to \emph{concurrently leverage all available data resources}, including limited labeled target data (\eg, labeled CT data), abundant unlabeled target data (\eg, unlabeled CT data) and well-established labeled source data (\eg, labeled MR data), to enhance the segmentation performance on target domain (\eg, CT).
We name our problem setting as semi-supervised domain adaptation (SSDA), following the previous work~\cite{saito2019semi}.

The straightforward SSDA approach is to jointly train labeled source data and labeled target data together, and then apply Pseudo-label method~\cite{lee2013pseudo} to tackle unlabeled target data.
Despite its simplicity for implementation, exploring all available data in one network exists several limitations.
First, the inter-domain knowledge acquired by joint training would be incompatible for target domain due to the apparent domain shift between source and target domains.
Moreover, low-quality pseudo labels would bring less accurate and biased training ground truth, greatly deteriorating intra-domain knowledge investigation.
Therefore, it would be beneficial to separately explore intra- and inter-domain features with specific networks, and then integrate dual-domain features for comprehensive integration.
In this paper, we propose a novel semi-supervised domain adaptation approach as a teacher-student framework, namely Dual-Teacher++.
Our entire framework consists of three components:
(1) an intra-domain teacher, which employs the self-ensembling model of the student network to leverage unlabeled target domain (\eg, CT) and transfers the acquired knowledge to student model by forcing prediction consistency;
(2) an inter-domain teacher, which adopts an image translation model, \ie, CycleGAN~\cite{zhu2017unpaired}, to narrow the appearance gap cross modalities and transfers the prior knowledge of source domain (\eg, MR) to student model via knowledge distillation;
and (3) a student model, which explicitly learns from limited labeled target data, and implicitly grasps auxiliary intra-domain and inter-domain knowledge transferred from two teachers for comprehensive integration and exploitation.

Furthermore, in the above framework, proper reliability control should be conducted to enhance the transferability of inter- and intra-domain knowledge.
For the inter-domain transfer, we adopt an image translation model (\eg, CycleGAN) to narrow down image-level appearance gap between source and target domains, while the GAN-based translator would synthesize some unrealistic target-domain-like images with poor quality at early training stage, bringing undesirable bias into the transfer.
Regarding the intra-domain transfer, since unlabeled target domain data has no expert-annotated labels, the intra-domain teacher model may produce unreliable and noisy predictions.
Transferring the knowledge beneath those predictions would be harmful for further exploitation.
To this end, we present specific reliability control strategies to promote reliable dual-domain knowledge transfer.
For the inter-domain transfer, we reweight each synthetic target-like sample with different importance via measuring its similarity to the target domain, and augment the transfer with high similarity.
When transferring the intra-domain knowledge, we calculate the prediction confidence of intra-domain teacher model via Monte Carlo sampling, and enhance the transfer with high confidence.
Overall, our framework is trained in an end-to-end manner to seamlessly integrate the latest dual-domain knowledge into the student model.

Our main contributions are summarized as follows.
\begin{itemize}
\item[(1)]  We propose a novel framework to extensively leverage auxiliary supervisions from cross-modality and unlabeled data for annotation-efficient cardiac segmentation.
We assign each data source with a specific network for non-inference exploration, and then transfer the acquired dual-domain knowledge into one network for further integration.

\item[(2)] To facilitate reliable knowledge transfer, we propose specific inter- and intra-domain reliability control strategies according to high confident synthetic target images and segmentation predictions, respectively.

\item[(3)] We extensively evaluated our method on the MM-WHS 2017 challenge dataset and our framework largely outperformed semi-supervised learning and domain adaptation methods.
%
We also conducted experiments on bidirectional cross-modality adaptation to further validate the superiority of our framework.
\end{itemize}

This work is a significant extension of our previous conference work Dual-Teacher~\cite{li2020dual}, regarding the following four aspects:
(1) we substantially improved our method with the dual-domain reliability control scheme, to mitigate the uncertain transfer and encourage reliable intra- and inter-domain knowledge integration. It brings significant performance gains over our previous work;
(2) We evaluated our framework with a larger amount of unlabeled data in training and more data for validation;
(3) we investigated bidirectional cross-modality adaptation from MR to CT and also from CT to MR, to further exhibit that our method could be well applied in both direction of domain adaptation, with no limitations on the choice of target domain;
and (4) we conducted more comprehensive comparisons with more state-of-the-art methods to show the feasibility of our method.

\section{Related works}
\begin{figure*}[!t]
\centering
\includegraphics[width=0.75\textwidth]{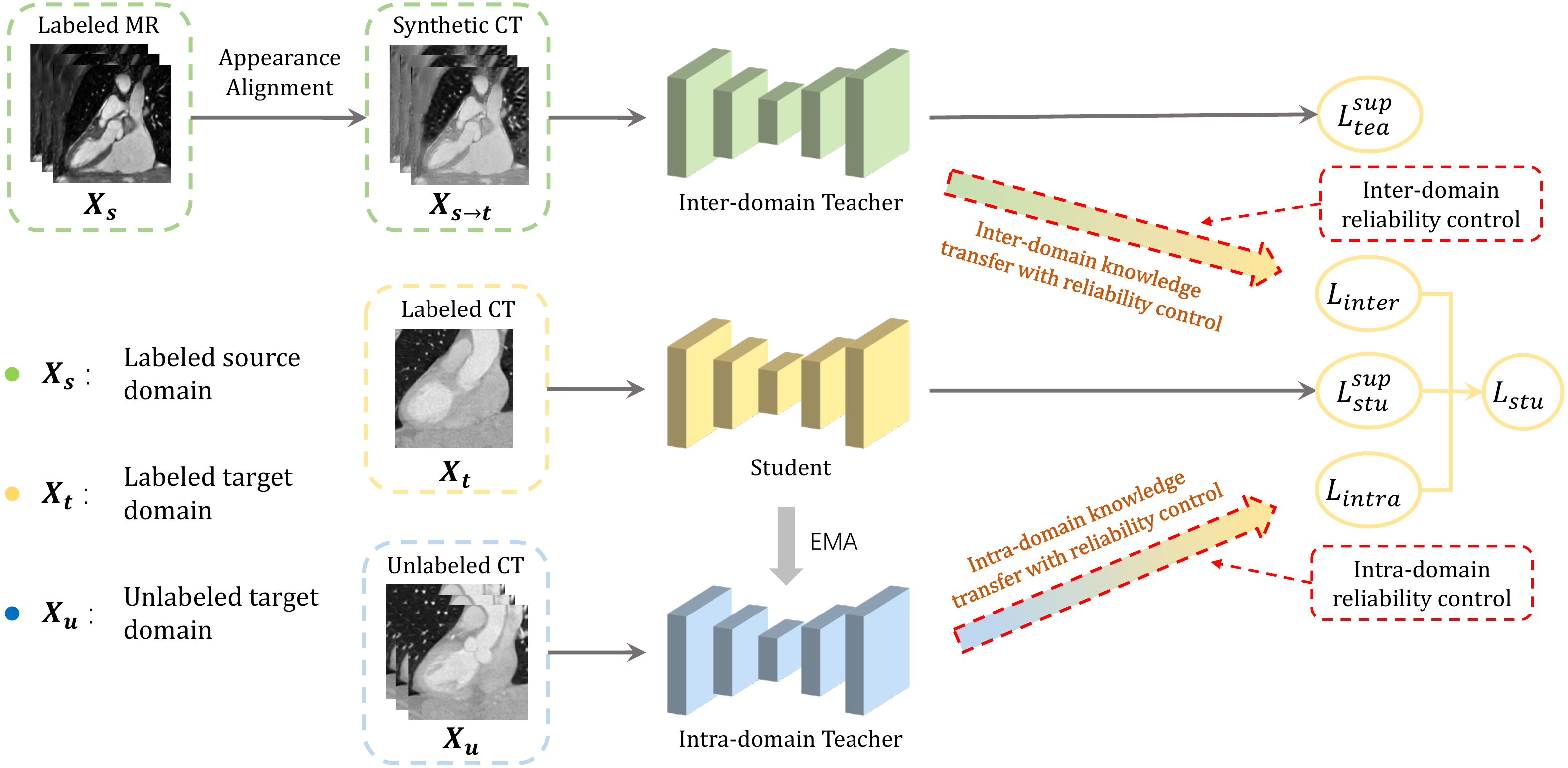}
\caption{Overview of our proposed Dual-Teacher++ framework. The student model learns directly from $\mathcal{X}_{t}$ via the $L_{stu}^{sup}$ loss, and concurrently acquires the knowledge in $\mathcal{X}_{s}$ from the inter-domain teacher with $L_{inter}$ loss and the knowledge in $\mathcal{X}_{u}$ from the intra-domain teacher via $L_{intra}$ loss.
To encourage reliable dual-domain knowledge transfer, an intra- and inter-domain reliability control scheme is conducted to facilitate further exploitation. For the inference, only the student model is used to predict.}
\label{fig:framework}
\end{figure*}
\subsection{Semi-supervised Learning}

Semi-supervised learning intends to leverage unlabeled data to alleviate annotation scarcity~\cite{lee2013pseudo,laine2016temporal,tarvainen2017mean,liu2016dual}.
Lee~\etal~\cite{lee2013pseudo} proposed to generate the pseudo labels of unlabeled data by a pretrained model, and utilized them to further finetune the training model for improved performance.
Recently, self-ensembling methods~\cite{laine2016temporal,tarvainen2017mean} have achieved state-of-the-art performance in many semi-supervised learning benchmarks.
Tarvainen~\etal~\cite{tarvainen2017mean} proposed the mean-teacher framework to force prediction consistency between the current training model and the exponential moving average (EMA) model.
Several approaches~\cite{li2020transformation,XIA2020101766,yu2019uncertainty} applied it in medical applications with further extensions like transformation-consistency constraints~\cite{li2020transformation}, multi-view co-training strategy~\cite{XIA2020101766} and uncertainty-aware consistency~\cite{yu2019uncertainty}.
Recent works~\cite{zhao2019data,chaitanya2019semi,frid2018gan} further adopted Generative Adversarial Networks~\cite{goodfellow2014generative} to synthesize realistic training examples for data augmentation and exploited them in semi-supervised approach.
However, semi-supervised learning approaches focus on exploiting the unlabeled data that belong to the same domain as labeled ones, leaving rich cross-modality data unexploited.

\subsection{Domain Adaptation}
Domain adaptation attempts to seek additional supervision from well-established cross-modality data, to promote the segmentation on target modality.
Many multi-modality learning approaches leveraged the modality-shared knowledge with specific feature fusion strategies, such as different parameter sharing strategies~\cite{van2018learning,valindria2018multi} and modality-specific normalization layers~\cite{dou2020unpaired}.
With the investigation of Generative Adversarial Network~\cite{goodfellow2014generative}, several works~\cite{cai2019towards,jue2019integrating} proposed to utilize GAN-based image translation model to align modality appearance first, and extract valuable modality-shared knowledge later.
However, MML approaches still require target modality annotations, while unsupervised domain adaptation extends it with a more annotation-efficient setting, where no target label is required.
Contemporary UDA methods attempt to extract domain-invariant representations, where Dou~\etal~\cite{dou2018unsupervised} investigated in feature space and Chen~\etal~\cite{chen2019synergistic} explored both feature-level and image-level in a synergistic manner.
However, there still exists considerable space for improvements.
Semi-supervised domain adaptation concentrates on an even more annotation-efficient setting, to concurrently leverage the additional supervisions from both cross-modality data and unlabeled data.
Very recently, several works tackled it with deep learning on computer vision applications~\cite{li2020online,saito2019semi,wang2019semi,wang2020alleviating,he2020classification}, however few efforts have been devoted in medical area.
Since cardiac segmentation annotations are often time-consuming and expensive to obtain, efficiently utilizing cardiac labels via semi-supervised domain adaptation becomes a promising yet under-explored branch.

\subsection{Knowledge Transfer}

Transfer learning aims at enhancing the model performance on target domain by transferring the valuable and helpful knowledge embedded in different but related source domains~\cite{zhuang2019comprehensive}.
Knowledge distillation was recently proposed to transfer knowledge from a cumbersome and deep model (\ie, teacher model) to a lightweight model (\ie, student model) for model compression~\cite{hinton2015distilling}.
Hinton~\etal~\cite{hinton2015distilling} stated that learning from teacher model's outputs would be more beneficial than directly learning from raw annotations, as the model outputs not only present which class is right to predict, but also reveal the inter-class similarity as soft targets~\cite{menon2020distillation,hinton2015distilling}.
It could be interpreted as a special format of label smoothing regularization by replacing the one-hot labels with smoothed ones~\cite{yuan2020revisiting}.
Several works~\cite{wang2019segmenting,kats2019soft,dou2020unpaired,li2020towards} have applied knowledge distillation into medical applications like cardiac segmentation~\cite{dou2020unpaired,li2020towards}.
Inspired by them, our framework applies knowledge distillation for knowledge transfer.
When transferring knowledge, one common concern is how to reliably transfer it.
Several works~\cite{su2020active,zhang2018importance,chen2020harmonizing,ding2020uncertainty,wang2019characterizing} proposed to assign each training sample with different importance, and the knowledge transfer of samples with larger importance would be enhanced to amplify reliable transfer.
Meanwhile, some approaches~\cite{yu2019uncertainty,XIA2020101766,sedai2019uncertainty} proposed to estimate model prediction confidence, where the regions with high confidence would be highlighted for reliable transfer.
Motivated by them, we also present specific reliability control in the light of intra- and inter-domain undependable cues to promote reliable transfer.

\begin{figure*}[!t]
\centering
\includegraphics[width=0.78\textwidth]{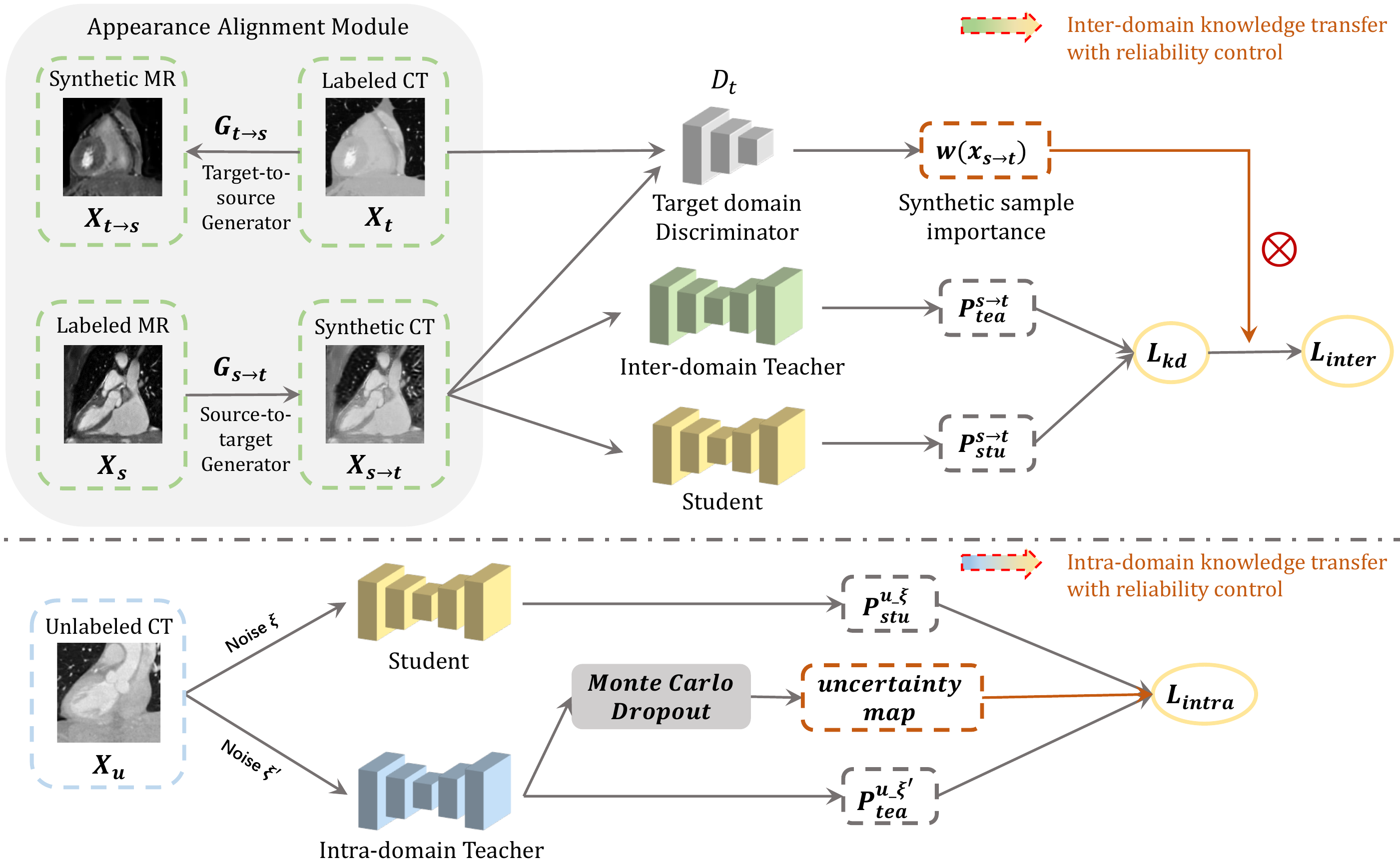}
\caption{The detailed diagram of inter-domain and intra-domain knowledge transfer. When transferring inter-domain knowledge, we encourage the student model prediction $P_{stu}^{s\rightarrow t}$ on synthetic target sample $x_{s \rightarrow t}$ to be similar to inter-domain teacher prediction $P_{tea}^{s\rightarrow t}$ via knowledge distillation. In addition, we promote reliable inter-domain knowledge transfer via assigning larger weights on the synthetic target samples which share more similarity to target domain. As for intra-domain knowledge transfer, given the same unlabeled data corrupted by small noises, we encourage the predictions of student model $P_{stu}^{u \_ \xi}$  to be consistent to those of intra-domain teacher model $P_{tea}^{u \_ \xi^{\prime}}$. Moreover, we facilitate reliable intra-domain knowledge transfer via suppressing the transfer with large prediction uncertainty. For simplicity, we omit the illustration of $D_{s}$ in appearance alignment module.}
\label{fig:inter-intra}
\end{figure*}

\section{Methodology}
In our problem setting, given a set of annotated images of source domain (\eg, labeled MR data) as  $\mathcal{X}_{s}=\left\{\left(\mathbf{x}_{i}^{s}, y_{i}^{s}\right)\right\}_{i=1}^{m_{s}}$, a limited number of annotated target domain samples (\eg, labeled CT data) as $\mathcal{X}_{t}=\left\{\left(\mathbf{x}_{i}^{t}, y_{i}^{t}\right)\right\}_{i=1}^{m_{t}}$, and abundant unlabeled target domain data (\eg, unlabeled CT data) as $\mathcal{X}_{u}=\left\{\left(\mathbf{x}_{i}^{u}\right)\right\}_{i=1}^{m_{u}}$.
Normally, we assume $m_{t}$ is far less than $m_{u}$.
Our goal is to exploit $\mathcal{X}_{s}, \mathcal{X}_{t}$, and $\mathcal{X}_{u}$ to enhance the performance in target domain (\eg, CT).
Fig.~\ref{fig:framework} overviews our Dual-Teacher++ framework, composed of two teacher models, one student model, and dual-domain reliability control.

\subsection{Inter-domain Teacher Model}

Source domain (\eg, MR) and target domain (\eg, CT) usually contain specific appearance and distinctive image distribution~\cite{pan2009survey, hoffman2018cycada}.
In this regard, we first narrow down the appearance differences to facilitate the inter-domain knowledge exploration,
and then exploit inter-domain reliability control to transfer dependable knowledge to student model.

\subsubsection{Inter-domain knowledge exploration}
Many domain adaptation methods~\cite{chen2019synergistic,li2020towards} proposed to eliminate the appearance discrepancy first before exploring modality-shared features.
We also adopt an appearance alignment module (AAM) to align source domain appearance to target domain.
Various image translation models can be utilized. Here we use CycleGAN~\cite{zhu2017unpaired} to translate source samples $x^{s}$ to synthetic target samples $x^{s \rightarrow t}$.
We stick to the original implementations and adopt the adversarial loss $\mathcal{L}_{adv}$  as:
\begin{equation}
\begin{aligned}
\mathcal{L}_{\text {adv}}^{t}(G_{s \rightarrow t}, D_{t})=
& \mathbb{E}_{x^{t} \sim X^{t}}[\log D_{t}(x^{t})]+ \\
& \mathbb{E}_{x^{s} \sim X^{s}}[\log (1-D_{t}(G_{s \rightarrow t}(x^{s})))], \\
\mathcal{L}_{\text {adv}}^{s}(G_{t \rightarrow s}, D_{s})=
& \mathbb{E}_{x^{s} \sim X^{s}}[\log D_{s}(x^{s})]+\\
& \mathbb{E}_{x^{t} \sim X^{t}}[\log (1-D_{s}(G_{t \rightarrow s}(x^{t})))], \\
\mathcal{L}_{adv}=\mathcal{L}_{\text {adv}}^{t}(
& G_{s \rightarrow t}, D_{t}) + \mathcal{L}_{\text {adv}}^{s}(G_{t \rightarrow s}, D_{s}),
\end{aligned}
\end{equation}
where $G_{s \rightarrow t}$ and $G_{t \rightarrow s}$ are the generators to perform source-to-target and target-to-source translation respectively, and $D_{s}$ and $D_{t}$ are the source domain discriminator and target domain discriminator, respectively.
In addition, we also impose cycle consistency constrain with $\mathcal{L}_{cyc}$ as
\begin{equation}
\begin{aligned}
    \mathcal{L}_{cyc}\left(G_{s \rightarrow t}, G_{t \rightarrow s}\right)=
    & \mathbb{E}_{x^{s} \sim X^{s}}\left\| \hat{x}^{s}-x^{s}\right \|_{1} + \\
    & \mathbb{E}_{x^{t} \sim X^{t}}\left\| \hat{x}^{t}-x^{t}\right \|_{1}.
\end{aligned}
\end{equation}
The full objective of appearance alignment module $\mathcal{L}_{AAM}$ is defined as
\begin{equation}
    \mathcal{L}_{AAM} = \mathcal{L}_{adv} + \lambda_{cyc}\mathcal{L}_{cyc},
\end{equation}
where $\lambda_{cyc}$ is a hyperparameter to control the relative importance of two objectives.
For simplicity, we omit the illustration of source domain discriminator $D_{s}$ in Fig.~\ref{fig:inter-intra}.

%
After appearance alignment, we feed synthetic samples $x^{s \rightarrow t}$ into the inter-domain teacher, which is implemented as a segmentation network. With the supervision of the corresponding labels $y^{s}$, the inter-domain teacher is able to learn the prior knowledge in source domain by $\mathcal{L}_{tea}^{seg}$.
Similar to previous works~\cite{milletari2016v, chen2019synergistic, yang2017hybrid, valindria2018multi}, we formulate it with a hybrid loss as
\begin{equation}
    \mathcal{L}_{tea}^{seg}=\mathcal{L}_{\mathrm{ce}}\left(y^{s}, p_{tea}^{s\rightarrow t}\right)+\mathcal{L}_{\mathrm{Dice}}\left(y^{s}, p_{tea}^{s\rightarrow t}\right),
\end{equation}
where $\mathcal{L}_{\mathrm{ce}}$ and $\mathcal{L}_{\mathrm{Dice}}$ denote cross-entropy loss and Dice loss, respectively, and $p_{tea}^{s\rightarrow t}$ represents the inter-domain teacher predictions taking $x^{s\rightarrow t}$ as inputs.
\subsubsection{Inter-domain reliable transfer}

To transfer inter-domain knowledge to the student model, we employ knowledge distillation~\cite{hinton2015distilling}, where the student model is encouraged to produce similar outputs as inter-domain teacher model, given the same synthetic sample $x^{s \rightarrow t}$.
Compared to directly feed synthetic target samples into the student model for inter-domain knowledge exploration, applying knowledge distillation via inter-domain teacher network is more informative~\cite{zhuang2019comprehensive,hinton2015distilling,menon2020distillation,yuan2020revisiting}.
%
It not only indicates the correct class to predict, but also reveals the inter-class correlation as soft labels.

However, not all synthetic target samples $x^{s \rightarrow t}$ are equally transferred.
Some synthetic samples may be more informative than others.
At the early training stage of CycleGAN, the generator $G_{s \rightarrow t}$ may lack the ability to produce satisfying results, but produce low-quality and less target-style synthetic samples.
Transferring the inter-domain knowledge of these samples may introduce unwanted bias and cause a negative impact on the student model.
To avoid negative transfer, we propose the inter-domain reliability control to reweight each synthetic target sample with different importance, as shown in Fig.~\ref{fig:inter-intra}.
We follow the idea of importance weighted empirical risk minimization (IWERM)~\cite{sugiyama2007covariate} to formulate the learning objective as a weighted knowledge distillation loss:
\begin{equation}
    \begin{aligned}
    \mathcal{L}_{inter}&=w(x^{s\rightarrow t})L_{kd} \\
        &=\frac{P_{T}(x)}{P_{S}(x)}\mathcal{L}_{\mathrm{ce}}\left(p_{tea}^{s \rightarrow t}, p_{stu}^{s \rightarrow t}\right),
\end{aligned}
\end{equation}
where we implement the knowledge distillation loss $L_{kd}$ as the cross-entropy loss following previous works~\cite{hinton2015distilling,anil2018large}, and $p_{tea}^{s \rightarrow t}$ and $p_{stu}^{s \rightarrow t}$ represent the predictions of inter-domain teacher model and student model, respectively.
Here $w(x^{s\rightarrow t})=\frac{p_{T}(x)}{p_{S}(x)}$ represents the importance of each synthetic target sample, and $P_{T}(x)$ and $P_{S}(x)$ denote the probability density function of target domain and source domain, respectively.

Intuitively, the synthetic samples that are of high quality and more similar to target domain are more reliable, and thus deserve higher importance during the transfer.
We take advantage of the target domain discriminator $D_{t}$ in appearance alignment module (i.e., CycleGAN) to determine the similarity of each synthetic target sample to real target domain data.
As $D_{t}$ is able to distinguish how realistic the current synthetic target samples are, its outputs would be a good indicator of the similarity to target domain. When $D_{t}$ arrives the optimal state~\cite{goodfellow2014generative}, we have
\begin{equation}
    w(x^{s \rightarrow t}) = \frac{D_{t}(x^{s \rightarrow t})}{1 - D_{t}(x^{s\rightarrow t})}.
    \label{eq: inter_weight}
\end{equation}
%
%

Assume we denote source domain as 0 and target domain as 1.
If the outputs of $D_{t}$ are closer to 1, it implies current synthetic samples are more similar to target domain and have fewer data uncertainty, and larger weights $w$ are given to them in transfer.
By substituting $w(x^{s \rightarrow t})$ by Eq.~\ref{eq: inter_weight}, our final objective function for inter-domain knowledge transfer is defined as
\begin{equation}
\mathcal{L}_{inter}=\frac{D_{t}(x^{s \rightarrow t})}{1 - D_{t}(x^{s\rightarrow t})}\mathcal{L}_{\mathrm{ce}}\left(p_{tea}^{s \rightarrow t}, p_{stu}^{s \rightarrow t}\right),
\label{eq: inter-trans}
\end{equation}
which indicates that (1) the data with higher empirical risk $L_{ce}$ (\ie, more inconsistent between the inter-teacher and student model outputs) is more important in optimization and (2) the sample with higher $D_{t}(x^{x\rightarrow t})$ outputs (\ie, more similar to target domain) is more crucial in training.

\subsection{Intra-domain Teacher Model}
Considering $\mathcal{X}_{u}$ has no expert-annotated labels, recent work~\cite{tarvainen2017mean} proposed to temporally ensemble the models for better guidance.
Inspired by it, we form intra-domain teacher model with the same network architecture as student model and update its weights $\theta^{\prime}$ as the exponential moving average (EMA) of the student model weights $\theta$ in different training steps.
At step $r$, the weights of intra-domain teacher model $\theta_{r}^{\prime}$ are updated as
\begin{equation}
\theta_{r}^{\prime}=\alpha \theta_{r-1}^{\prime}+(1-\alpha) \theta_{r},
\end{equation}
where $\alpha$ is the EMA decay rate to control the updating rate.
%

Intuitively, to explore and transfer intra-domain knowledge to student model, we would force prediction consistency regularization.
To be specific, we encourage the student model to generate the same outputs as the corresponding EMA model (\ie, intra-domain teacher model), given the same unlabeled data under different perturbations.
However, with the absence of ground truth for unlabeled data, intra-domain teacher model would inevitably produce unreliable or uncertain predictions.
Thus we present the intra-domain reliability control to promote the transfer of regions with high prediction confidence.
Following Gal~\etal~\cite{gal2016dropout} and others~\cite{yu2019uncertainty,XIA2020101766,sedai2019uncertainty}, we estimate the prediction uncertainty by Monte Carlo Dropout with Bayesian networks.
To be specific, we input a batch of unlabeled target data $x^{u}$ with random dropout and different Gaussian noise for $N$ stochastic forward passes through intra-domain teacher model.
By doing so, we would obtain a set of probability maps $\left\{\mathbf{p}_{i}\right\}_{i=1}^{N}$.
Following Kendall~\etal~\cite{kendall2017bayesian}, we utilize entropy to estimate the uncertainty, since it has a fixed range, where the max uncertainty is supposed to be $ln(C)$, where $C$ represents the total number of classes to predict.
The mean prediction $\mu_{c}$ and uncertainty $u$ at each pixel are estimated as
\begin{equation}
    \begin{aligned}
    \mu_{c}&=\frac{1}{N} \sum_{i=1}^{N} \mathbf{p}_{i}^{c}, \\
    u&=-\sum_{c=1}^{C} \mu_{c} \log \mu_{c},
\end{aligned}
\end{equation}
where $\mathbf{p}_{i}^{c}$ denotes the probability map of the $c$-th class in the $i$-th stochastic forward pass.
Note that the above uncertainty is estimated in pixel level, and the whole image uncertainty $U=\{u\} \in \mathbb R^{H \times W}$, where $H$ and $W$ denote for image height and width, respectively.
As the predictive entropy as a fixed range~\cite{gal2016dropout}, we use an indicator function to filter out unreliable predictions in uncertainty map, and only transfer the knowledge from the regions with high confidence to student model.
The final objective for intra-domain knowledge transfer $\mathcal{L}_{intra}$  with reliability control is defined as
\begin{equation}
    \begin{aligned}
        \mathcal{L}_{con}(f,f^{\prime})=\mathcal{L}_{\mathrm{mse}}\left(f\left(x^{u};\theta,\xi\right), f\left(x^{u};\theta^{\prime},\xi^{\prime}\right)\right),\\
        \mathcal{L}_{intra} =\frac{\sum_{i=1}^{W}\sum_{j=1}^{H} \mathbb{I}\left(u_{i,j}< u_{thre}\right)\mathcal{L}_{con}(f,f^{\prime})}{\sum_{i=1}^{W}\sum_{j=1}^{H} \mathbb{I}\left(u_{i,j}<u_{thre}\right)},
    \label{eq: intra-trans}
    \end{aligned}
\end{equation}
where $\mathcal{L}_{\mathrm{mse}}$ denotes mean squared error loss.  $f\left(x^{u};\theta,\xi\right)$ and $f\left(x^{u};\theta^{\prime},\xi^{\prime}\right)$ represent the outputs of student model (with weight $\theta$ and noise $\xi$) and intra-domain teacher model (with weight $\theta^{\prime}$ and noise $\xi^{\prime}$), respectively.
$\mathbb{I}(\cdot)$ is the indicator function, $u_{i,j}$ is the estimated uncertainty at position $(i, j)$, and $u_{thre}$ is a threshold to filter out the pixels with high uncertainty.
Assisted by intra-domain reliability control, student model would integrate dependable intra-knowledge for exploitation.

\subsection{Student Model and Overall Training Strategies}
In our framework, the student model first explicitly learns from $\mathcal{X}_{t}$ supervised by corresponding annotations via the segmentation loss $\mathcal{L}_{stu}^{seg}$.
Meanwhile, the student model also concurrently acquires the knowledge of $\mathcal{X}_{s}$ and $\mathcal{X}_{u}$ from inter- and intra-domain teacher models under dual-domain reliable control, to comprehensively integrate them as a united cohort.
The training objective for student model $\mathcal{L}_{stu}$ is formulated as
\begin{equation}
\begin{aligned}
\mathcal{L}_{stu}^{seg}&=\mathcal{L}_{\mathrm{ce}}\left(y^{t}, p_{stu}^{t}\right)+\mathcal{L}_{\mathrm{Dice}}\left(y^{t}, p_{stu}^{t}\right),\\
\mathcal{L}_{stu}&=\mathcal{L}_{stu}^{seg} + \lambda_{kd}\mathcal{L}_{inter} + \lambda_{con}\mathcal{L}_{intra},
\end{aligned}
\label{eq: whole-obj-function}
\end{equation}
where $\lambda_{kd}$ and $\lambda_{con}$ are hyperparameters for the tradeoff of $\mathcal{L}_{inter}$ and $\mathcal{L}_{intra}$.
%
Our framework is updated in an end-to-end manner. We first optimize inter-domain teacher model and the student model, and update intra-domain teacher model with the EMA parameters of the student network.

\section{Experiments}

\begin{table*}[th]
\centering
\caption{Performance comparisons with other methods on CT cardiac segmentation. The Dice and ASD of all heart substructures and the average of them are reported here. We have highlighted the best results in \textbf{bold}.}

    \begin{tabular}{c|c|C{1.2cm}|C{1cm}|C{1cm}|C{1cm}|C{1cm}|C{1cm}|C{1cm}|C{1cm}}
    \toprule[1pt]
    \multicolumn{2}{c|}{\multirow{2}{*}{Method}} & \multirow{2}{*}{Avg $\uparrow$ } & \multicolumn{7}{c}{Dice $[\%]$ of heart substructures $\uparrow$ } \\ \cline{4-10}
    \multicolumn{2}{c|}{} & & MYO & LA & LV & RA & RV & AA & PA\\
    \midrule
    \multicolumn{2}{c|}{Supervised-only ($\mathcal{X}_{t}$)} & 72.75 & 71.81 & 75.35 & 78.20 &  71.89 & 68.24 & 82.13 & 61.65 \\ \hline

    \multirow{2}{*}{\begin{tabular}[c]{@{}c@{}}UDA\\ $(\mathcal{X}_{s}, \mathcal{X}_{u})$\end{tabular}} & Dou~\etal\cite{dou2018unsupervised} & 66.56 & 59.15 & 81.38 & 76.59 & 57.75 & 65.97 & 64.20 & 44.91\\
      & Chen~\etal\cite{chen2019synergistic} & 72.26 & 70.64 & 83.39 & 84.30 & 80.99 & 74.88 & 73.10 & 38.56 \\ \hline

    \multirow{6}{*}{\begin{tabular}[c]{@{}c@{}}MML\\ $(\mathcal{X}_{s}, \mathcal{X}_{t})$\end{tabular} }
    & Finetune & 74.22 & 72.31 & 80.87 & 82.91 & 74.18 & 67.40 & 84.75 & 57.12  \\
      & Joint training & 78.37 & 79.45 & 85.29 & 87.44 & 75.57 & 63.55 & 88.55 & 68.78 \\
      & X-shape~\cite{valindria2018multi} & 76.44 & 71.62 & 85.26 & 84.01 & 68.94 & 70.18 & 88.16 & 66.93 \\
      & Dou~\etal~\cite{dou2020unpaired} & 81.66 & 76.81 & 82.96 & 88.27 & 81.03 & 81.44 & 85.41 & 75.68 \\
      & Cai~\etal~\cite{cai2019towards} & 80.95 & 81.46 & 81.97 & 88.88 & 79.66 & 75.16 & 88.62 & 70.91  \\
      & MKD~\cite{li2020towards} & 82.33 & 81.55 & 83.96 & 90.22 & 80.29 & 79.55 & 92.67 & 68.12 \\
      \hline

      \multirow{2}{*}{\begin{tabular}[c]{@{}c@{}}SSL\\ $(\mathcal{X}_{u}, \mathcal{X}_{t})$\end{tabular}}
      & MT~\cite{tarvainen2017mean} & 82.73 & 80.65 & 86.43 & 88.85 & 81.59 & 74.49 & 90.89 & 76.22 \\
      & UA-MT~\cite{yu2019uncertainty} & 83.22 & 79.50 & 88.25 & 88.22 & 82.13 & 70.98 & 92.20 & 81.30 \\ \hline

    \multirow{2}{*}{\begin{tabular}[c]{@{}c@{}}SSDA\\ $(\mathcal{X}_{s}, \mathcal{X}_{u}, \mathcal{X}_{t})$\end{tabular}}
    & Dual-Teacher~\cite{li2020dual} & 86.44 & 84.96 & \textbf{89.54} & \textbf{92.71} & 85.12 & 76.53 & \textbf{95.25} & 80.95  \\
    & \textbf{Dual-Teacher++ (Ours)} & \textbf{87.82} & \textbf{85.20} & 89.26 & 92.07 & \textbf{85.18} & \textbf{84.46} & 95.06 & \textbf{83.54} \\

    \bottomrule[1pt]
    \end{tabular}
    \\[4pt]
    \begin{tabular}{c|c|C{1.2cm}|C{1cm}|C{1cm}|C{1cm}|C{1cm}|C{1cm}|C{1cm}|C{1cm}}
    \toprule[1pt]
    \multicolumn{2}{c|}{\multirow{2}{*}{Method}} & \multirow{2}{*}{Avg $\downarrow$ } & \multicolumn{7}{c}{ASD [voxel] of heart substructures $\downarrow$} \\ \cline{4-10}
    \multicolumn{2}{c|}{} & & MYO & LA & LV & RA & RV & AA & PA\\
    \midrule
    \multicolumn{2}{c|}{Supervised-only ($\mathcal{X}_{t}$)} & 10.13 & 5.31 & 7.64 & 5.91 & 17.88 & 12.84 & 7.80 & 13.55 \\ \hline

    \multirow{2}{*}{\begin{tabular}[c]{@{}c@{}}UDA\\ $(\mathcal{X}_{s}, \mathcal{X}_{u})$\end{tabular}}
    & Dou~\etal\cite{dou2018unsupervised} & 12.71 & 10.04 & 8.06 & 11.80 & 14.31 & 16.81 & 13.06 & 14.92 \\
    & Chen~\etal\cite{chen2019synergistic} & 10.77 & 7.61 & 6.77 & 8.61 & 11.42 & 13.68 & 11.21 & 16.07 \\ \hline

    \multirow{6}{*}{\begin{tabular}[c]{@{}c@{}}MML\\ $(\mathcal{X}_{s}, \mathcal{X}_{t})$\end{tabular} }
    & Finetune & 10.20 & 5.95 & 8.42 & 9.13 & 12.11 & 14.70 & 6.68 & 14.43 \\
      & Joint training & 8.04 & 4.03 & 6.54 & 3.91 & 10.72 & 15.20 & 3.94 & 11.91 \\
      & X-shape~\cite{valindria2018multi} & 8.32 & 6.81 & 6.32 & 8.17 & 12.96 & 7.60 & 4.11 & 12.28 \\
      & Dou~\etal~\cite{dou2020unpaired} & 7.18 & 4.78 & 7.34 & 4.18 & 8.14 & 8.57 & 6.14 & 11.13 \\
      & Cai~\etal~\cite{cai2019towards} & 6.83 & 2.87 & 8.19 & 3.82 & 12.22 & 6.50 & 3.76 & 10.45 \\
      & MKD~\cite{li2020towards} & 6.68 & 2.93 & 6.31 & 3.54 & 8.60 & 9.42 & 5.26 & 10.72 \\
      \hline

      \multirow{2}{*}{\begin{tabular}[c]{@{}c@{}}SSL\\ $(\mathcal{X}_{u}, \mathcal{X}_{t})$\end{tabular}}
      & MT~\cite{tarvainen2017mean} & 6.26 & 3.49 & 5.36 & 3.55 & 8.49 & 6.61 & 4.88 & 11.46 \\
      & UA-MT~\cite{yu2019uncertainty} & 5.11 & 3.70 & 5.25 & 2.57 & 7.17 & 6.20 & 3.35 & 7.52 \\ \hline

    \multirow{2}{*}{\begin{tabular}[c]{@{}c@{}}SSDA\\ $(\mathcal{X}_{s}, \mathcal{X}_{u}, \mathcal{X}_{t})$\end{tabular}}
    & Dual-Teacher~\cite{li2020dual} & 4.51 & 3.34 & \textbf{4.11} & 2.63 & \textbf{5.31} & 5.75 & 2.67 & 7.74 \\
    & \textbf{Dual-Teacher++ (Ours)} & \textbf{3.69} & \textbf{2.81} & 4.42 & \textbf{2.04} & 6.08 & \textbf{4.50} & \textbf{1.44} & \textbf{4.53} \\
    \bottomrule[1pt]
    \end{tabular}
\label{tab: ct-dice-asd-main}
\end{table*}

\begin{figure*}[th]
\centering
\includegraphics[width=0.75\textwidth]{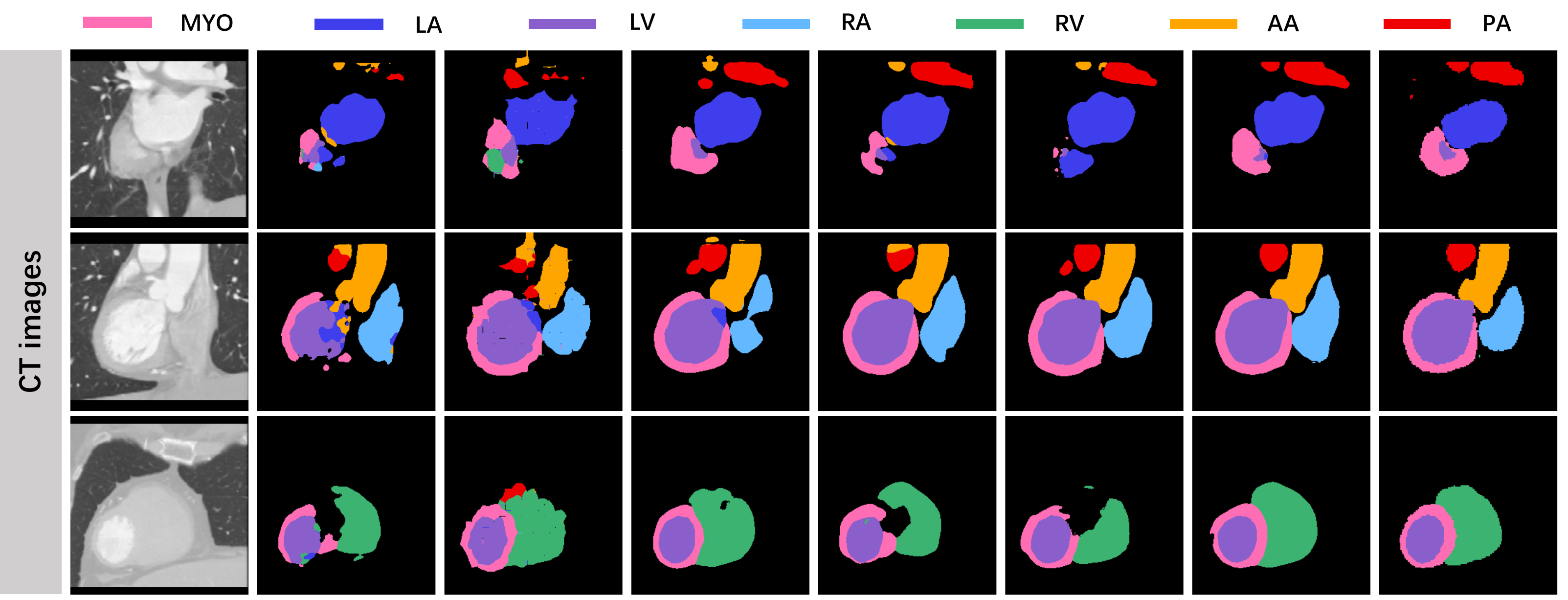}
\includegraphics[width=0.75\textwidth]{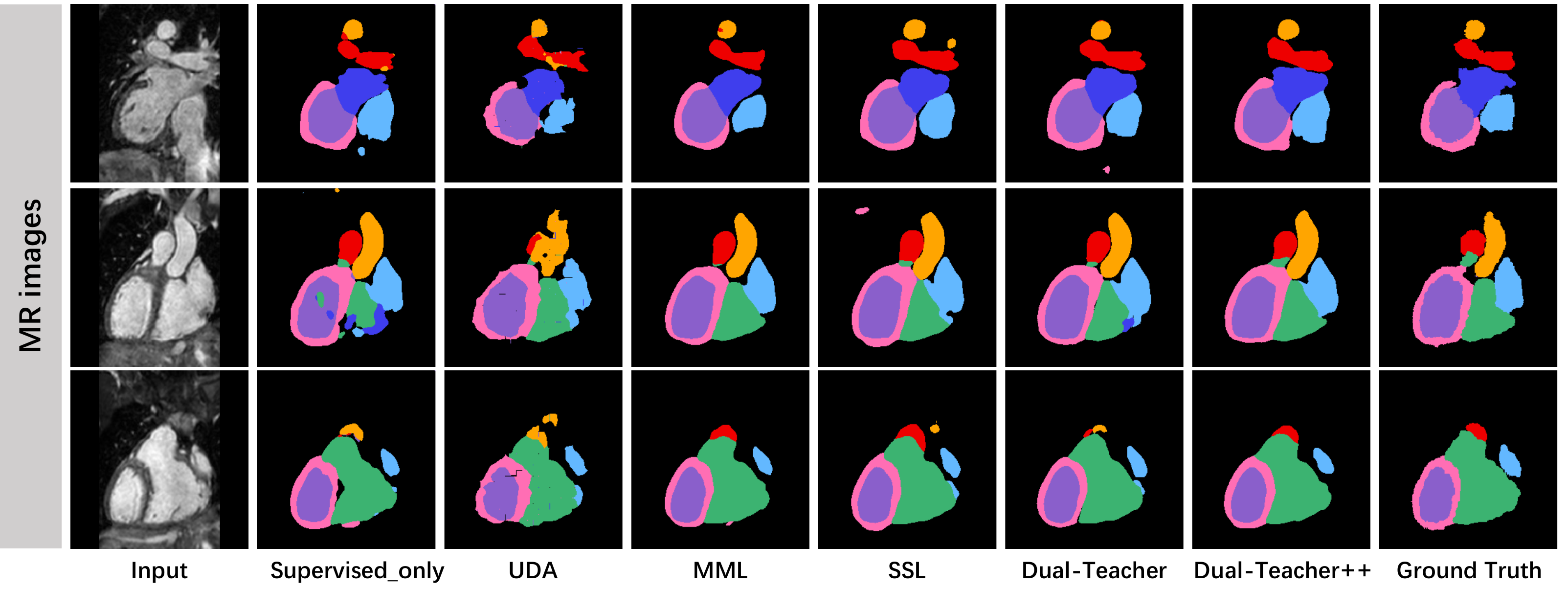}
\caption{Visual Comparisons on cardiac segmentation. Here we only present the methods with best mean Dice in UDA, MML and SSL (\ie, Chen~\etal~\cite{chen2019synergistic}, MKD~\cite{li2020towards} and UA-MT~\cite{yu2019uncertainty}). As observed, our predictions are more similar to the ground truth than others.}
\label{fig:ct-seg}
\end{figure*}

\subsection{Dataset and Pre-processing}

We extensively evaluated our framework on MM-WHS 2017 dataset, which provides 20 labeled and 40 unlabeled CT volumes, as well as 20 labeled and 40 unlabeled MR volumes.
We evaluated our method with bidirectional cross-modality domain adaptation, \ie, from CT to MR and from MR to CT. In each direction of adaptation, we performed four-fold cross-validation.
Take the adaptation from MR to CT as an example.
We randomly split 20 annotated CT volumes into four-fold.
In each fold of validation, we took one fold (\ie, 5 volumes) CT data for training as labeled target domain $\mathcal{X}_{t}$ and the remaining three folds (\ie, 15 volumes) CT data for testing.
We also adopted all 20 labeled MR data as labeled source domain $\mathcal{X}_{s}$ and 40 unlabeled CT data as unlabeled target domain $\mathcal{X}_{u}$ for training.
The adaptation from CT to MR applies to the same data splitting rule.

For pre-processing, we resampled all volumes with unit spacing and cropped them centering at the heart region, following previous work~\cite{chen2019synergistic}.
We also utilized on-the-fly data augmentation with random affine transformations and random rotation to avoid overfitting.
We evaluated our method with both Dice coefficient [$\%$] and Average Surface Distance (ASD) [voxel] on all heart substructures, including the left ventricle blood cavity (LV), the  right  ventricle  blood  cavity  (RV), the left atrium  blood  cavity  (LA), the right atrium blood cavity (RA), the myocardium of the left ventricle (MYO), the ascending aeorta (AA), and the pulmonary artery (PA)~\cite{zhuang2019evaluation}.
Dice coefficient measures the overlap ratio between the predictions and the ground truth, while ASD assesses the average distances between the surface of predictions and the ground truth with the unit of voxel.
Thus, a higher Dice value and a lower ASD result would imply better segmentation performance.

\subsection{Implementation Details}
All components in our framework are online updated in an end-to-end manner. For each iteration, we start with the optimization of appearance alignment module, then update inter-domain teacher, student model, and intra-domain teacher in order.
%
In our framework, we implemented the student model and two teacher models with the same framework \ie, U-Net~\cite{ronneberger2015unet}.
%
%
Following~\cite{yu2019uncertainty}, we empirically set EMA decay rate $\alpha$ as $0.99$ and temporally changed the hyperparapmeter $\lambda_{con}$ with function $\lambda_{con}(t)=0.1 * e^{\left(-5\left(1-t / t_{\max }\right)^{2}\right)}$, where $t$ and $t_{max}$ represent for current training epoch and the last epoch. Here, we set $t_{max}$ as 50.
To adapt the intra-domain teacher model as Bayesian network for uncertainty estimation, we added two dropout layers with 0.25 dropout rate at the convolution layer with the most condensed features (i.e., the last convolution layer in the contracting path) and the last convolution layer before the final prediction in U-Net.
Following the previous work~\cite{yu2019uncertainty}, we empirically set $N=8$ to balance the uncertainty quality and training speed, and ramped up the uncertainty threshold similarly as EMA decay rate.
Similar to Yu~\etal~\cite{yu2019uncertainty}, the threshold {$u_{thre}$} gradually varies from $\frac{3}{4}u_{max}$ to $u_{max}$, where $u_{max}=\ln{8}$ in our case.
For inter-domain teacher model, we empirically set $\lambda_{kd}$ as 5 to balance the importance between inter- and intra-domain knowledge transfer.
The optimizer settings of appearance alignment module are closely followed CycleGAN~\cite{zhu2017unpaired}.
We employed Adam optimizer with learning rate 0.0001 to optimize model parameters until convergence.

\subsection{Comparison with Other Methods}
To validate the feasibility of our Dual-Teacher++ in leveraging extra supervisions of unlabeled data $\mathcal{X}_{u}$ and cross-modality data $\mathcal{X}_{s}$, we comprehensively compared with both semi-supervised learning methods and domain adaptation methods, as shown in TABLE~\ref{tab: ct-dice-asd-main}.
We first trained a model with only labeled target domain data $\mathcal{X}_{t}$ and referred it as Supervised-only.
For semi-supervised learning, we compared our method with mean-teacher~\cite{tarvainen2017mean}, which has demonstrated its effectiveness at many computer vision benchmarks.
We further compared with UA-MT~\cite{yu2019uncertainty}, which also employed self-ensembling model on cardiac segmentation.
For domain adaptation, we compared our framework with both unsupervised domain adaptation approach and multi-modality learning approach.
We took two UDA methods for comparisons, \ie, Dou~\etal~\cite{dou2018unsupervised} and Chen~\etal~\cite{chen2019synergistic}, which achieved the cutting-edge performance in cardiac segmentation.
For MML methods, we compared with several straightforward ones like finetune and joint training, and other advanced methods with different strategies to leverage modality-shared knowledge, including X-shape~\cite{valindria2018multi} via parameter sharing strategies, Dou~\etal~\cite{dou2020unpaired} via knowledge distillation, Li~\etal~\cite{li2020towards} by mutual learning and also Cai~\etal~\cite{cai2019towards} via appearance alignment model.

Table~\ref{tab: ct-dice-asd-main} reports the cardiac segmentation performance from MR (\ie, source domain) to CT (\ie, target domain) evaluated by both Dice and ASD.
Trained only with labeled CT data, the Supervised-only achieved $72.75\%$ in mean Dice and 10.13 voxel error in average ASD.
With the help of unlabeled CT data, both semi-supervised learning methods including MT and UA-MT achieved significant performance gains.
The best of them (\ie, UA-MT~\cite{yu2019uncertainty}) would increase the mean Dice with $10.47\%$ gain and decrease the mean ASD to 5.02.
When cross-modality data is available for investigation, the UDA method like Chen~\etal~\cite{chen2019synergistic} was able to obtain considerable performance as Supervised-only, even when no target annotation is provided, showing the feasibility of using cross-modality prior knowledge.
With labeled cross-modality data $\mathcal{X}_{s}$ and limited labeled target modality data $\mathcal{X}_{t}$, all MML methods could achieve improved performance over Supervised-only. The best of them (\ie, MKD~\cite{li2020towards}) yielded $9.58\%$ improvements in mean Dice and reduced mean ASD with 3.45 voxels, further validating the effectiveness of cross-modality knowledge.
With our proposed Dual-Teacher++ framework, both unlabeled data $\mathcal{X}_{t}$ and cross-modality data $\mathcal{X}_{s}$ could be well exploited, leading to $4.60\%$, $5.49\%$, $15.56\%$, $15.07\%$ enhancements in mean Dice and 1.42, 2.99, 7.08 and 6.44 decrease in mean ASD, compared to the best method in SSL, MML, UDA, and Supervised-only, respectively.
The dual-domain reliability control would provide additional guidance to facilitate reliable inter- and intra- knowledge transfer, and thus yielded $1.38\%$ improvements in mean Dice than our previous work Dual-Teacher.
We also provide visual comparisons in Fig.~\ref{fig:ct-seg}. As observed, our framework could better identify multi-substructure with less false positive predictions, especially for PA and RV. Moreover, our method could produce more accurate segmentation maps with clean and smooth boundaries.
%
%
\begin{table*}[!th]
\centering
\caption{
Analysis of key components in our method. We report the mean Dice and mean ASD of all cardiac substructures, and highlight the best results in \textbf{bold}.}

%
\begin{tabular}{c|C{3cm}|C{1.5cm}|C{1.5cm}|C{2cm}|C{2cm}|C{1.3cm}|C{1.3cm}}
\toprule[1pt]
\multicolumn{2}{c|}{Methods}                                                             & Intra-domain teacher & Inter-domain teacher & Intra-domain reliability control & Inter-domain reliability control & Mean Dice $[\%]$ $\uparrow$ & Mean ASD [voxel] $\downarrow$ \\
\midrule
\multirow{2}{*}{No-Teacher}
& Baseline  &  &  &   &   & 74.49 & 9.76 \\ \cline{2-8}
& GAN-baseline  &   &    &   &  & 77.85 & 9.15 \\ \hline
\multicolumn{1}{l|}{\multirow{2}{*}{One-Teacher}}
& W/o intra-domain teacher  &   & $\checkmark$   &    &    & 80.48 & 7.32   \\ \cline{2-8}
\multicolumn{1}{c|}{}
& W/o inter-domain teacher  & $\checkmark$ &  &  &  & 85.03 & 5.26  \\ \hline
\multirow{4}{*}{Two-Teacher}
& Vanilla two-teacher model   &  $\checkmark$ &  $\checkmark$   &  &      & 86.44 & 4.51 \\ \cline{2-8}
& W/o intra-domain reliability control &  $\checkmark$ &  $\checkmark$ &  &  $\checkmark$ & 86.85 &  4.38 \\ \cline{2-8}
& W/o inter-domain reliability control &  $\checkmark$  &  $\checkmark$  &  $\checkmark$ &    & 87.22  & 3.84  \\ \cline{2-8} & \textbf{Dual-Teacher++ (Ours)}  & $\checkmark$  & $\checkmark$  & $\checkmark$  & $\checkmark$ & \textbf{87.82} & \textbf{3.69} \\
\bottomrule[1pt]
\end{tabular}
\label{tab:ablation}
\end{table*}

\begin{table*}[!th]
\centering
\caption{Performance comparisons with other methods on MR cardiac segmentation, where the Dice and ASD of all heart substructures and the average of them are reported here. The highest Dice and lowest ASD are highlighted in \textbf{bold}.}

    \begin{tabular}{c|c|C{1.2cm}|C{1cm}|C{1cm}|C{1cm}|C{1cm}|C{1cm}|C{1cm}|C{1cm}}
    \toprule[1pt]
    \multicolumn{2}{c|}{\multirow{2}{*}{Method}} & \multirow{2}{*}{Avg $\uparrow$ } & \multicolumn{7}{c}{Dice $[\%]$ of heart substructures $\uparrow$} \\ \cline{4-10}
    \multicolumn{2}{c|}{} & & MYO & LA & LV & RA & RV & AA & PA\\
    \midrule
    \multicolumn{2}{c|}{Supervised-only ($\mathcal{X}_{t}$)} & 67.06 & 67.56 & 68.34 & 81.89 & 71.90 & 69.01 & 67.61 & 43.09 \\ \hline

    \multirow{2}{*}{\begin{tabular}[c]{@{}c@{}}UDA\\ $(\mathcal{X}_{s}, \mathcal{X}_{u})$\end{tabular}}
    & Dou~\etal\cite{dou2018unsupervised} & 58.77
 & 55.47 & 53.57 & 76.89 & 77.78 & 60.05 & 52.93 & 34.72 \\
    & Chen~\etal\cite{chen2019synergistic} & 62.31 & 60.03 & 55.92 & 80.39 & 79.97 & 59.40 & 59.16 & 41.33 \\ \hline

    \multirow{6}{*}{\begin{tabular}[c]{@{}c@{}}MML\\ $(\mathcal{X}_{s}, \mathcal{X}_{t})$\end{tabular} }
    & Finetune & 71.49 & 70.07 & 67.45 & 84.53 & 75.79 & 70.18 & 68.69 & 63.72  \\
      & Joint training & 74.97 & 71.57 & 72.99 & 84.59 & 81.50 & 74.10 & 72.66 & 67.38 \\
      & X-shape~\cite{valindria2018multi} & 72.14 & 69.30 & 69.20 & 82.77 & 78.07 & 64.47 & 71.35 & 69.78 \\
      & Dou~\etal~\cite{dou2020unpaired} & 80.46 & 76.31 & 80.83 & 90.57 & 84.80 & 79.07 & 77.02 & 74.64 \\
      & Cai~\etal~\cite{cai2019towards} & 79.61 & 75.31 & 79.62 & 88.58 & 85.65 & 80.26 & 76.01 & 71.82 \\
      & MKD~\cite{li2020towards} & 80.91 & 77.98 & 82.14 & 89.82 & 87.55 & 82.12 & 75.15 & 71.63 \\
      \hline

      \multirow{2}{*}{\begin{tabular}[c]{@{}c@{}}SSL\\ $(\mathcal{X}_{u}, \mathcal{X}_{t})$\end{tabular}}
      & MT~\cite{tarvainen2017mean} & 76.11 & 71.17 & 76.35 & 85.53 & 79.11 & 74.09 & 76.32 & 70.16 \\
      & UA-MT~\cite{yu2019uncertainty} & 77.70 & 74.27 & 77.09 & 89.02 & 80.87 & 77.36 & 75.28 & 70.03 \\ \hline

    \multirow{2}{*}{\begin{tabular}[c]{@{}c@{}}SSDA\\ $(\mathcal{X}_{s}, \mathcal{X}_{u}, \mathcal{X}_{t})$\end{tabular}}
    & Dual-Teacher~\cite{li2020dual} & 83.04 & 77.54 & \textbf{85.18} & 91.82 & \textbf{88.97} & 85.13 & 78.60 & 74.06 \\
    & \textbf{Dual-Teacher++ (Ours)} & \textbf{84.81} & \textbf{82.48} & 84.70 & \textbf{93.34} & 86.71 & \textbf{90.10} & \textbf{79.28} & \textbf{77.09} \\

    \bottomrule[1pt]
    \end{tabular}
\\[4pt]
\begin{tabular}{c|c|C{1.2cm}|C{1cm}|C{1cm}|C{1cm}|C{1cm}|C{1cm}|C{1cm}|C{1cm}}
    \toprule[1pt]
    \multicolumn{2}{c|}{\multirow{2}{*}{Method}} & \multirow{2}{*}{Avg $\downarrow$ } & \multicolumn{7}{c}{ASD [voxel] of heart substructures $\downarrow$} \\ \cline{4-10}
    \multicolumn{2}{c|}{} & & MYO & LA & LV & RA & RV & AA & PA\\
    \midrule
    \multicolumn{2}{c|}{Supervised-only ($\mathcal{X}_{t}$)} & 9.49 & 9.13 & 8.91 & 4.55 & 10.02 & 12.04 & 10.31 & 11.48 \\ \hline

    \multirow{2}{*}{\begin{tabular}[c]{@{}c@{}}UDA\\ $(\mathcal{X}_{s}, \mathcal{X}_{u})$\end{tabular}}
    & Dou~\etal\cite{dou2018unsupervised} & 11.55 & 13.17 & 11.92 & 5.44 & 9.40 & 13.12 & 12.08 & 15.73 \\
    & Chen~\etal\cite{chen2019synergistic} & 10.54 & 10.14 & 9.61 & 4.42 & 8.86 & 14.03 & 14.06 & 12.65  \\ \hline

    \multirow{6}{*}{\begin{tabular}[c]{@{}c@{}}MML\\ $(\mathcal{X}_{s}, \mathcal{X}_{t})$\end{tabular} }
    & Finetune & 8.09 & 5.36 & 8.65 & 4.24 & 9.61 & 9.68 & 9.42 & 9.70 \\
      & Joint training & 7.13 & 5.07 & 7.10 & 4.33 & 6.02 & 7.35 & 9.24 & 10.81 \\
      & X-shape~\cite{valindria2018multi} & 7.65 & 4.21 & 6.88 & 4.93 & 6.53 & 10.74 & 9.20 & 11.04 \\
      & Dou~\etal~\cite{dou2020unpaired} & 6.21 & 4.81 & 7.02 & 3.84 & 6.37 & 8.61 & 4.66 & 8.13 \\
      & Cai~\etal~\cite{cai2019towards} & 6.65 & 5.34 & 6.91 & 5.24 & 6.17 & 8.90 & 4.48 & 9.51  \\
      & MKD~\cite{li2020towards} & 5.84 & 4.75 & 4.26 & 4.22 & 4.84 & 7.82 & 5.44 & 9.55  \\
      \hline

      \multirow{2}{*}{\begin{tabular}[c]{@{}c@{}}SSL\\ $(\mathcal{X}_{u}, \mathcal{X}_{t})$\end{tabular}}
      & MT~\cite{tarvainen2017mean} & 6.37 & 5.71 & 5.92 & 5.37 & 6.28 & 6.41 & 5.14 & 9.79 \\
      & UA-MT~\cite{yu2019uncertainty} & 5.67 & 4.14 & 5.64 & 2.85 & 6.46 & 5.62 & 5.86 & 9.13  \\ \hline

    \multirow{2}{*}{\begin{tabular}[c]{@{}c@{}}SSDA\\ $(\mathcal{X}_{s}, \mathcal{X}_{u}, \mathcal{X}_{t})$\end{tabular}}
    & Dual-Teacher ~\cite{li2020dual} & 5.31 & 5.16 & \textbf{4.13} & \textbf{2.56} & \textbf{3.94} & 7.18 & 5.23 & 8.94 \\
    & Dual-Teacher++ (Ours) & \textbf{4.49} & \textbf{3.42} & 4.55 & 3.31 & 5.24 & \textbf{5.56} & \textbf{4.05} & \textbf{5.31} \\

    \bottomrule[1pt]
    \end{tabular}
\label{tab: mr-dice-asd-main}
\end{table*}

\begin{table*}[!th]
\centering
\caption{Paired sample t-test between Dual-Teacher and Dual-Teacher++. The p-values lower than the significance level, i.e., 0.05, are highlighted in \textbf{bold}.}

\begin{tabular}{c|c|c|c|c|c|c|c|c|c}
\toprule[1pt]
                          & \begin{tabular}[c]{@{}l@{}}Evaluation \\ Metric\end{tabular} & Avg & MYO & LA & LV & RA & RV & AA & PA \\
\midrule
\multirow{2}{*}{MR to CT} & Dice & \bm{$2.36 e^{-4}$} & $7.15 e^{-1}$ & $5.28 e^{-1}$ & $3.44 e^{-1}$ & $9.56 e^{-1}$ & \bm{$2.29 e^{-5}$} & $4.98 e^{-1}$ & \bm{$3.19 e^{-2}$} \\
                          & ASD  & \bm{$4.49 e^{-2}$} & $4.72 e^{-1}$ & \bm{$7.15 e^{-3}$} & $1.55 e^{-1}$ & \bm{$1.17 e^{-2}$} & \bm{$1.11 e^{-3}$} & \bm{$6.43 e^{-4}$} & \bm{$3.18 e^{-3}$} \\
\hline
\multirow{2}{*}{CT to MR} & Dice & \bm{$6.18 e^{-5}$} & \bm{$1.42 e^{-3}$} & $5.38 e^{-1}$ & \bm{$2.36 e^{-2}$} & \bm{$1.95 e^{-2}$} & \bm{$4.29 e^{-4}$} & $5.32 e^{-1}$ & \bm{$3.58 e^{-3}$} \\
                          & ASD  & \bm{$2.70 e^{-2}$} & \bm{$1.57 e^{-3}$} & $5.17 e^{-1}$ & $1.76 e^{-1}$ & \bm{$2.90 e^{-3}$} & \bm{$1.46 e^{-2}$} & \bm{$2.52 e^{-2}$} & \bm{$1.18 e^{-5}$} \\
\bottomrule[1pt]
\end{tabular}
\label{tab: paired-t-test}
\end{table*}

\begin{figure*}[]
\centering
\includegraphics[width=0.98\textwidth]{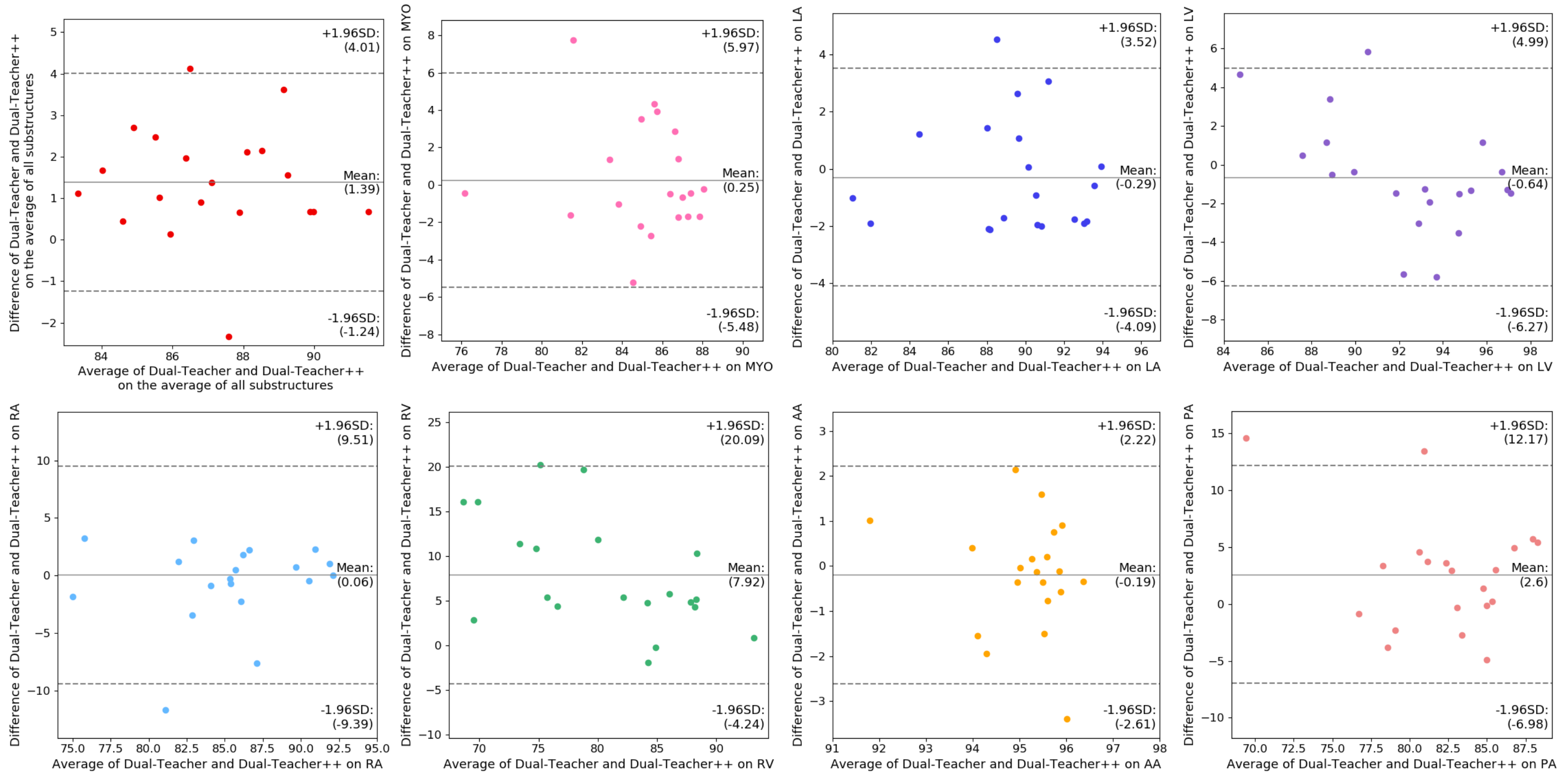}
\includegraphics[width=0.98\textwidth]{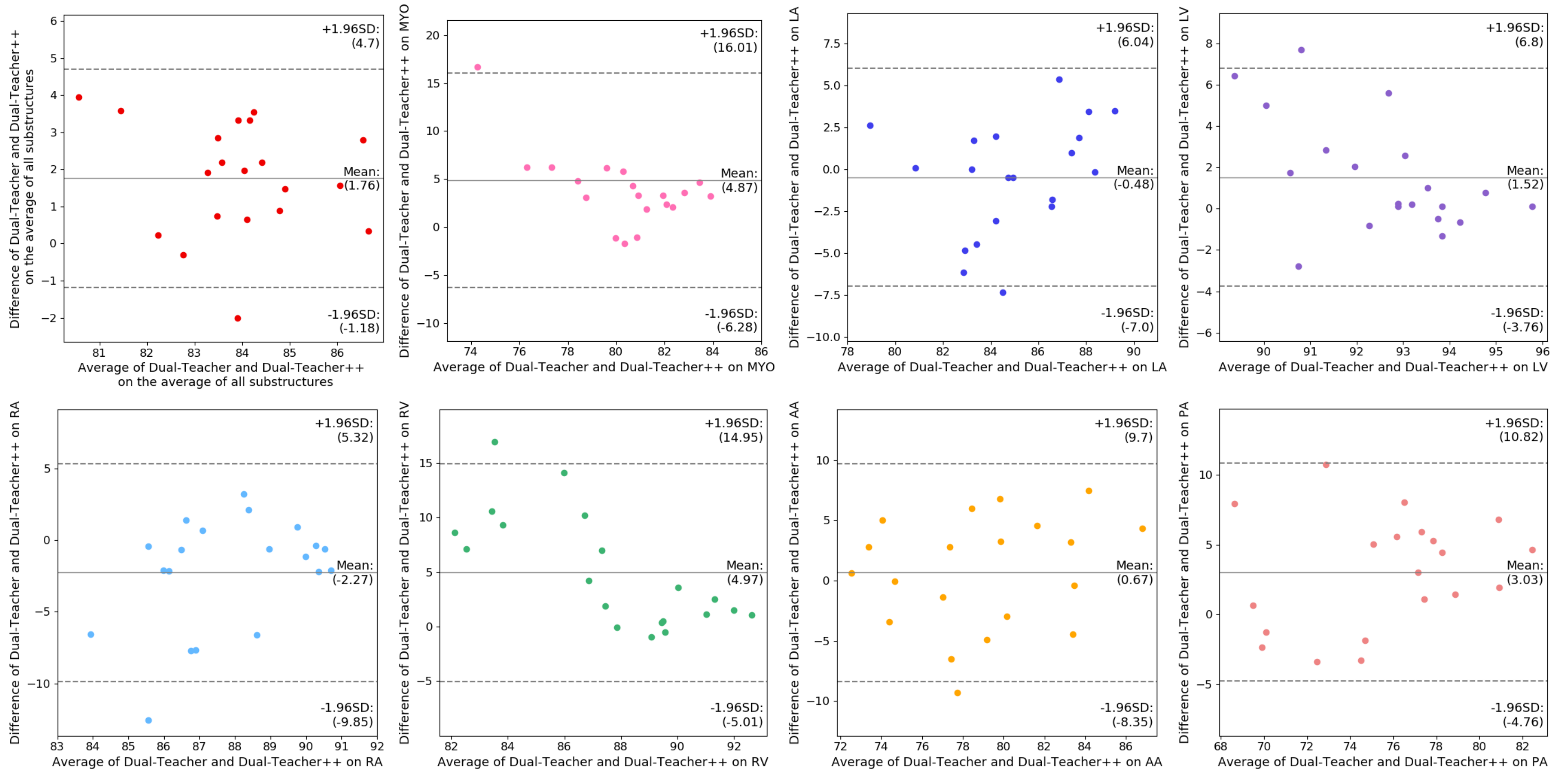}
\caption{Bland-Altman plots between Dual-Teacher and Dual-Teacher++ of Dice results for bidirectional domain adaptation, where the first two rows and the last two rows present the adaptation from MR to CT and from CT to MR, respectively.}
\label{fig: Bland-Altman}
\end{figure*}

\subsection{Ablation Analysis}

Table~\ref{tab:ablation} presents the ablation studies of our key components. We first implemented the straightforward SSDA method via joint training cross-modality data $\mathcal{X}_{s}$ with limited labeled target data $\mathcal{X}_{t}$, and exploring abundant unlabeled data $\mathcal{X}_{u}$ by Pseudo-label approach~\cite{lee2013pseudo}, which we referred as baseline.
A more effective baseline (referred as GAN-baseline) is to employ GAN-based image translation model to align cross-modality appearance first, and then follow the previous routine in baseline.
In the implementations of baseline and GAN-baseline, no teacher scheme is adopted.
With the help of GAN-based image translation model, the eliminated appearance differences reduced the difficulty in modality-shared knowledge exploration, leading to improved performance.
However, as no teaching scheme is applied, there exist no intra- or inter- knowledge transfer and neither cross-modality data nor unlabeled data would be well-exploited.
The performance of both baselines exist considerable spaces for improvements.
%

%
To analyze the effectiveness of our proposed dual-teaching scheme, we put our dual-domain reliability control aside, and focus on the proposed teacher-student framework for the moment, namely Vanilla two-teacher model.
The vanilla two-teacher model setting is the same as our prior Dual-teacher.
%
When only one teacher is presented, without intra-domain teacher model, we replaced it with Pseudo-label approach to investigate unlabeled data $\mathcal{X}_{u}$.
With the absence of inter-domain teacher model, we jointly trained $\mathcal{X}_{s}$ and $\mathcal{X}_{t}$ with appearance alignment module attached.
Without intra-domain teacher model, less accurate pseudo labels would cause biased training targets.
The bias directly interference the optimization of all model parameters in the current iteration, and the consequences continue affecting the following iterations through gradient back-propagation, resulting in degraded segmentation performance with $5.96\%$ lower than Vanilla two-teacher model in mean Dice.
When inter-domain teacher model was absent, joint training was not effective enough to investigate the knowledge beneath cross-modality data.
Joint training encourages the student to directly explore inter-domain knowledge from raw annotations.
Compared with exploring them by inter-domain teacher via knowledge distillation, it is less informative as it only indicates the correct class to predict, but ignores the inter-class correlation, deteriorating the segmentation performance with $1.41\%$ decrease in mean Dice.

We further experimented on the effectiveness of the proposed dual-domain reliability transfer.
We further conducted experiments when only one reliability control was applied, \ie, w/o intra-domain reliability control and w/o inter-domain reliability control, to observe separate performance gains of them.
As observed in Table~\ref{tab:ablation}, either intra- or inter- domain reliability control would guide more reliable domain knowledge integrated into the student model, and thus yielded extra performance gains compared to vanilla two-teacher model.
With Dual-domain reliability control presented, both intra- and inter- domain reliability would be well considered, and achieved $1.38\%$ improvements than vanilla two-teacher model.

\subsection{Bidirectional Knowledge Transfer}

To further exhibit that our performance gains are bidirectional, we switched the roles of CT and MR by taking CT as source domain while MR as target domain.
Table~\ref{tab: mr-dice-asd-main} presents MR cardiac segmentation performance evaluated by Dice and ASD, respectively.
Trained with limited labeled MR data $\mathcal{X}_{t}$, the Supervised-only achieved $67.06\%$ in mean Dice and 9.49 in mean ASD.
By further exploiting well-established cross-modality data, the best multi-modality learning method would greatly enhance the performance with $13.85\%$ in average Dice and $3.65$ in average ASD, respectively.
When a large quantity of unlabeled data is available, the best semi-supervised learning method improved MR segmentation results with $10.64\%$ in mean Dice and $3.82$ in average ASD.
As our proposed framework could sufficiently leverage cross-modality data and unlabeled data, and reliably integrate them, our Dual-Teacher++ yielded $3.90\%$ higher mean Dice and 1.35 lower mean ASD over MML, and $7.11\%$ higher mean Dice and 1.18 lower mean ASD over SSL, indicating the effectiveness of our method.
With the help of dual-domain reliability control, our Dual-Teacher++ further achieved $1.77\%$ enhancement in mean Dice, compared with prior Dual-Teacher.
We also provided visual comparisons on MR segmentation predictions in Fig.~\ref{fig:ct-seg}. As observed, our method produced more similar predictions to ground truth over other methods, and the predicted contours are smoother.
The above experiments validated that our performance gains are bidirectional.
In either the adaptation from MR to CT or CT to MR, our method would exhibit superior performance with significant improvements over others.

To better show the significance of Dual-Teacher++ over our previous work, we conducted a paired sample T-Test between Dual-Teacher++ and Dual-Teacher in Table IV and produced Bland-Altman Plots of each cardiac substructures in Fig.~\ref{fig: Bland-Altman}.
%
Here we set the significance level as $5 e^{-2}$.
As observed, the improvements over our previous work are significant in the average of substructures under either direction of adaptation.
In particular, when adapting from MR to CT, for several substructures that are hard to segment due to low image contrast and ambiguous boundaries, like RV and PA, our enhancements are clearly significant.
Meanwhile, in the adaptation from CT to MR, the p-values of most heart substructures like MYO, RA, RV and PA are lower than our significance level, indicating significant promotion.

\section{Discussion}
Annotation scarcity has been a crucial problem in deep learning based methods.
This situation is even more severe in medical area, where labeling them has high demands on clinical expertise.
Many approaches proposed to tackle it by resorting to other available data resources, where semi-supervised learning sought to abundant unlabeled data and domain adaptation leveraged widely available cross-modality data.
The effectiveness of each one of them has been extensively validated in many medical applications. However, few studies concentrate on the feasibility of leveraging the merits of them both.
Our work proceeds along this promising direction and demonstrates that concurrently exploiting unlabeled data and cross modality data would yield much more performance gains than only investigating one of them.
Compared with prior Dual-Teacher, our Dual-Teacher++ further proposed dual-domain reliability control to promote reliable knowledge transfer.
Equipped with more reliable intra- and inter-domain knowledge, our framework outperformed prior Dual-Teacher on the segmentation of several substructures including MYO, RV and PA.
On the contrary, prior Dual-Teacher may easily misidentify PA with AA, and MYO with LA in CT segmentation, as shown in the first two rows in Fig.~\ref{fig:ct-seg}.
Prior Dual-Teacher may also produce incomplete predictions on RV (the third row in Fig.~\ref{fig:ct-seg}).

Considering the trade-off between computation efficiency and accuracy, the proposed Dual-Teacher++ utilized 2D networks to perform volumetric cardiac segmentation.
Our entire framework is composed of multiple components and trained in an end-to-end manner.
The appearance alignment module (\ie, CycleGAN), two teacher models and one student model are optimized together to synchronously integrate the latest dual-domain knowledge into the student model.
Implementing our framework in 3D is complicated and more importantly memory intensive, with much more computation and training time cost. We intend to solve it in our future work.

One promising yet challenging extension of our work is to further investigate the knowledge of unlabeled cross-modality data.
Given only limited labels of target domain, this new setting will exploit the knowledge from all available data resources within the same label space. It has none limitations on data modality or annotations.
Kalluri~\etal~\cite{kalluri2019universal} named this setting as universal semi-supervised domain adaptation and proposed a model with pixel-level entropy regularization to jointly connect target domain and cross-modality datasets.
It enlightens another promising direction towards an even more annotation-efficient setting for cardiac segmentation.

\section{Conclusion}
This paper presents a novel semi-supervised domain adaptation framework, namely Dual-Teacher++, for annotation-efficient cardiac segmentation.
We employ two teacher models to investigate inter- and intra-domain knowledge and transfer them into student model under dual-domain reliability control for further integration and exploitation.
We extensively validate our method on MM-WHS 2017 dataset with bidirectional cross-modality adaptation from MR to CT and from CT to MR.
The experimental results demonstrate the effectiveness of our framework in terms of both Dice and ASD value, compared with other semi-supervised learning methods and domain adaptation methods.

\bibliographystyle{IEEEtran}
\bibliography{ref}

\end{document}